\documentclass[epj,twocolumn]{webofc}
\usepackage[varg]{txfonts}   
\usepackage{adjustbox}
\usepackage{url}
\usepackage{import}
\usepackage{xcolor}
\woctitle{Powders \& Grains 2017}
\begin{document}
\title{Pomelo, a tool for computing Generic Set Voronoi Diagrams of Aspherical Particles of Arbitrary Shape}
%
%

\author{\firstname{Simon} \lastname{Weis}\inst{1}\fnsep\thanks{\email{simon.weis@fau.de}} \and
        \firstname{Philipp W.~A.} \lastname{Schönhöfer}\inst{1, 2}\fnsep\thanks{\email{P.Schoenhoefer@murdoch.edu.au} \newline SW and PS have contributed equally to the work described in this project} \and
        \firstname{Fabian M.} \lastname{Schaller}\inst{1}\and 
        \firstname{Matthias} \lastname{Schröter}\inst{3}\and
        \firstname{Gerd E.} \lastname{Schröder-Turk}\inst{1, 2}
}

\institute{Institut für Theoretische Physik I, Universität Erlangen-Nürnberg, Staudtstraße 7, 91058 Erlangen, Germany
\and
           School of Engineering and Information Technology, Murdoch University, 90 South Street, Murdoch, WA 6150, Australia
\and
           Institute for Multiscale Simulation, Nägelsbachstrasse 49b, 91052 Erlangen, Germany 
          }

\abstract{%
We describe the development of a new software tool, called "Pomelo", for the calculation of Set Voronoi diagrams. 
Voronoi diagrams are a spatial partition of the space around the particles into separate Voronoi cells, e.g. applicable to granular materials. 
A generalization of the conventional Voronoi diagram for points or monodisperse spheres is the Set Voronoi diagram, also known as navigational map or tessellation by zone of influence. 
In this construction, a Set Voronoi cell contains the volume that is closer to the surface of one particle than to the surface of any other particle. 
This is required for aspherical or polydisperse systems.

Pomelo is designed to be easy to use and as generic as possible. 
It directly supports common particle shapes and offers a generic mode, which allows to deal with any type of particles that can be described mathematically. 
Pomelo can create output in different standard formats, which allows direct visualization and further processing.
Finally, we describe three applications of the Set Voronoi code in granular and soft matter physics, namely the problem of packings of ellipsoidal particles with varying degrees of particle-particle friction, mechanical stable packings of tetrahedra and a model for liquid crystal systems of particles with shapes reminiscent of pears. 
}
\maketitle
%


\begin{figure*}[hbtp]
    \centering
    \begin{adjustbox}{minipage=0.25\linewidth}
    \def\svgwidth{\linewidth}
    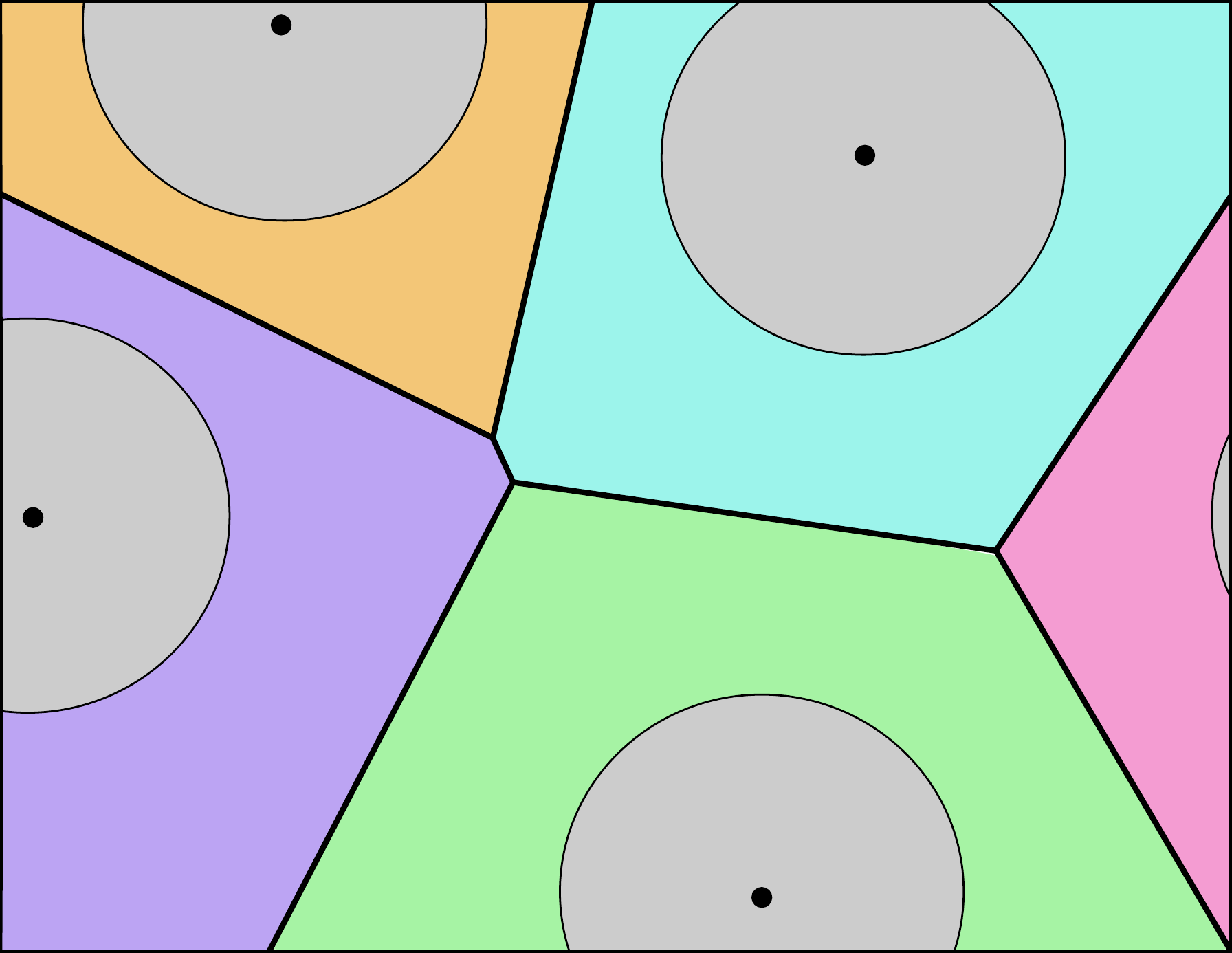
    \end{adjustbox}
    \begin{adjustbox}{minipage=0.25\linewidth}
    \def\svgwidth{\linewidth}
    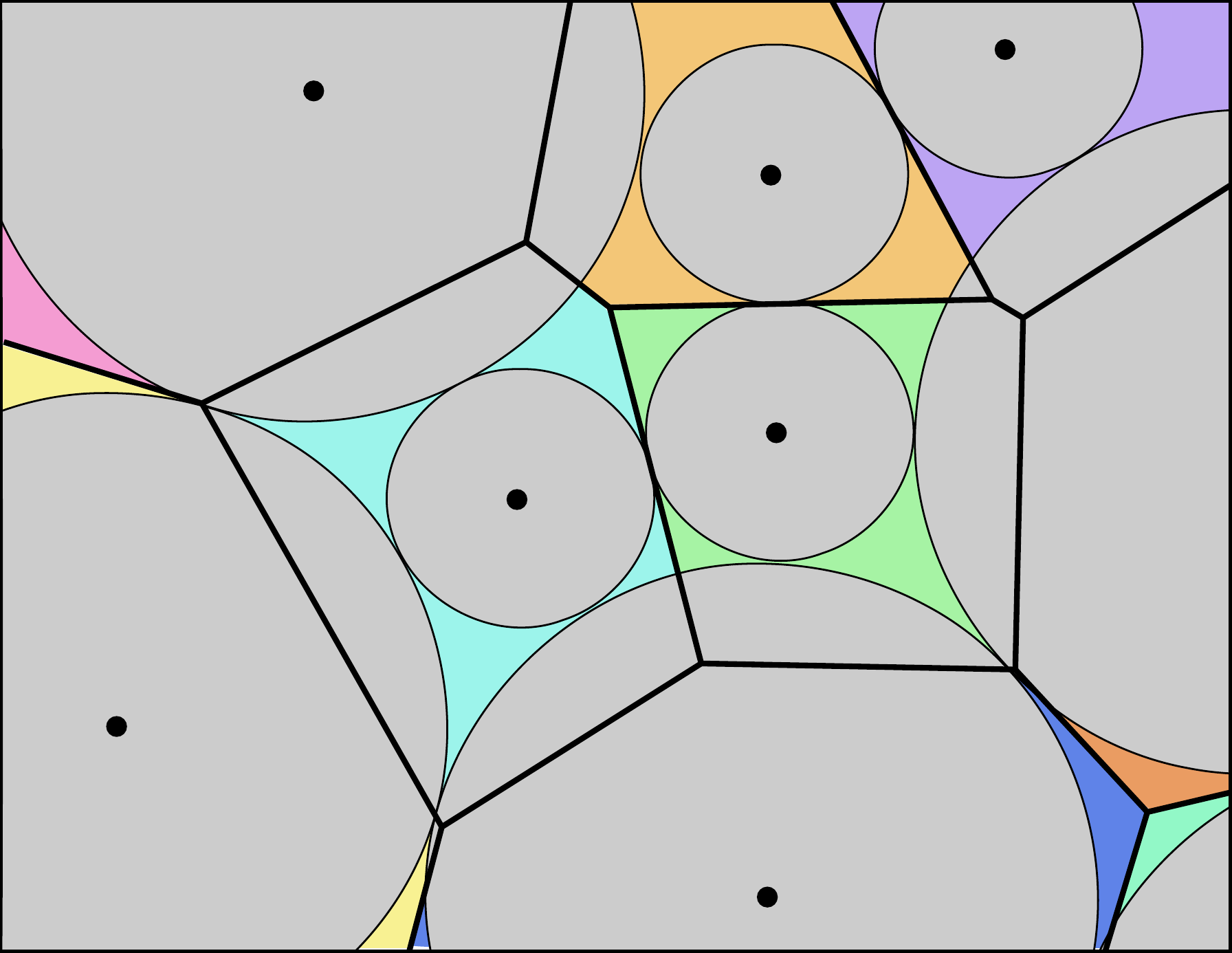
    \end{adjustbox}
    \begin{adjustbox}{minipage=0.25\textwidth}
    \def\svgwidth{\linewidth}
    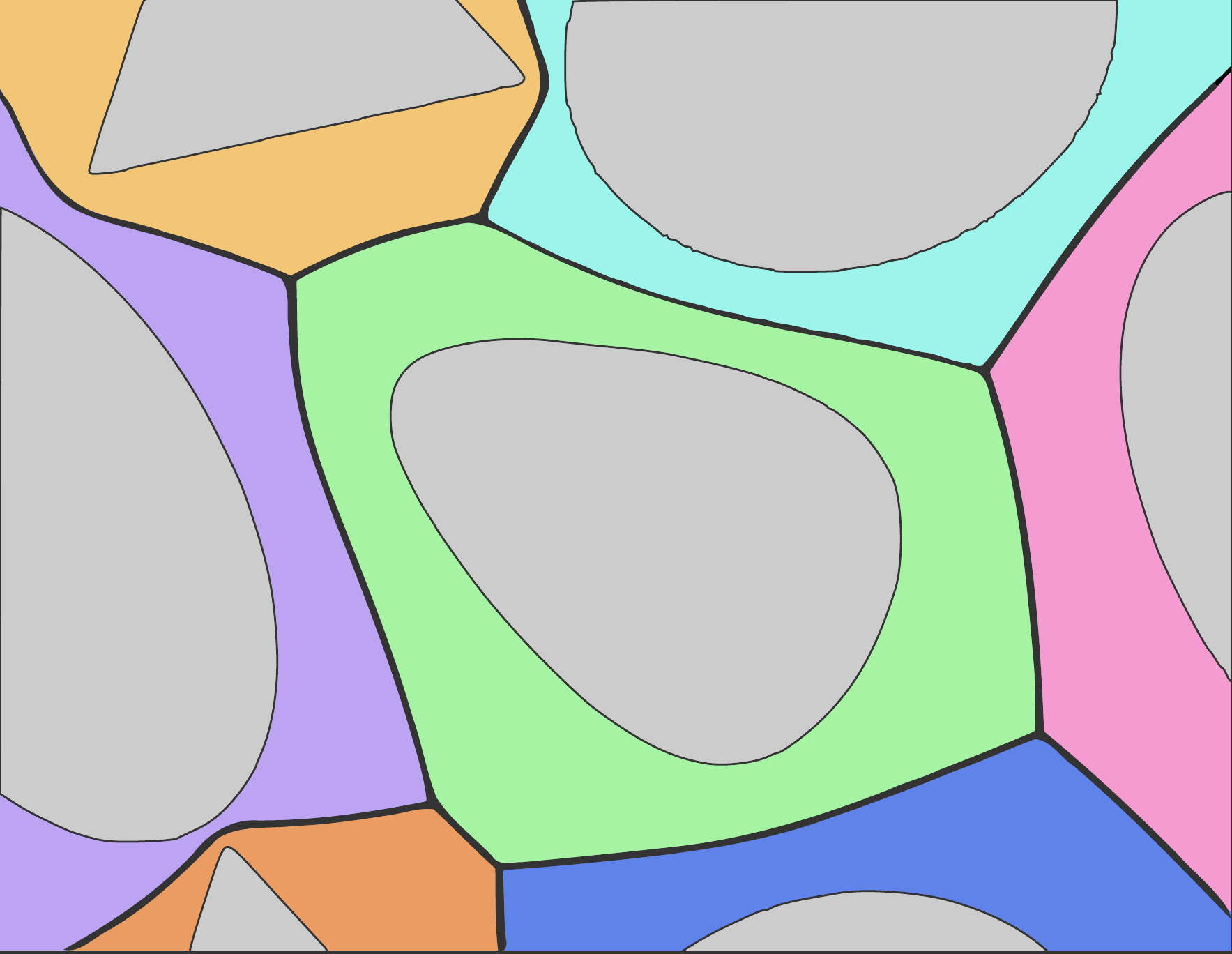
    \end{adjustbox}
    \caption{Left: The Voronoi Diagram of a system of monodisperse spheres. Center: The Voronoi Diagram is not suitable for a system of bidisperse spheres, as Voronoi cells are overlapping with particles. Right: The Set Voronoi Diagram of a mixture of differently shaped objects.}
    \label{fig:voronoi-setvoronoi}
\end{figure*}

\begin{figure*}[htbp]
    \centering
    \hfill
    \begin{adjustbox}{minipage=0.19\textwidth}
    \def\svgwidth{\textwidth}
    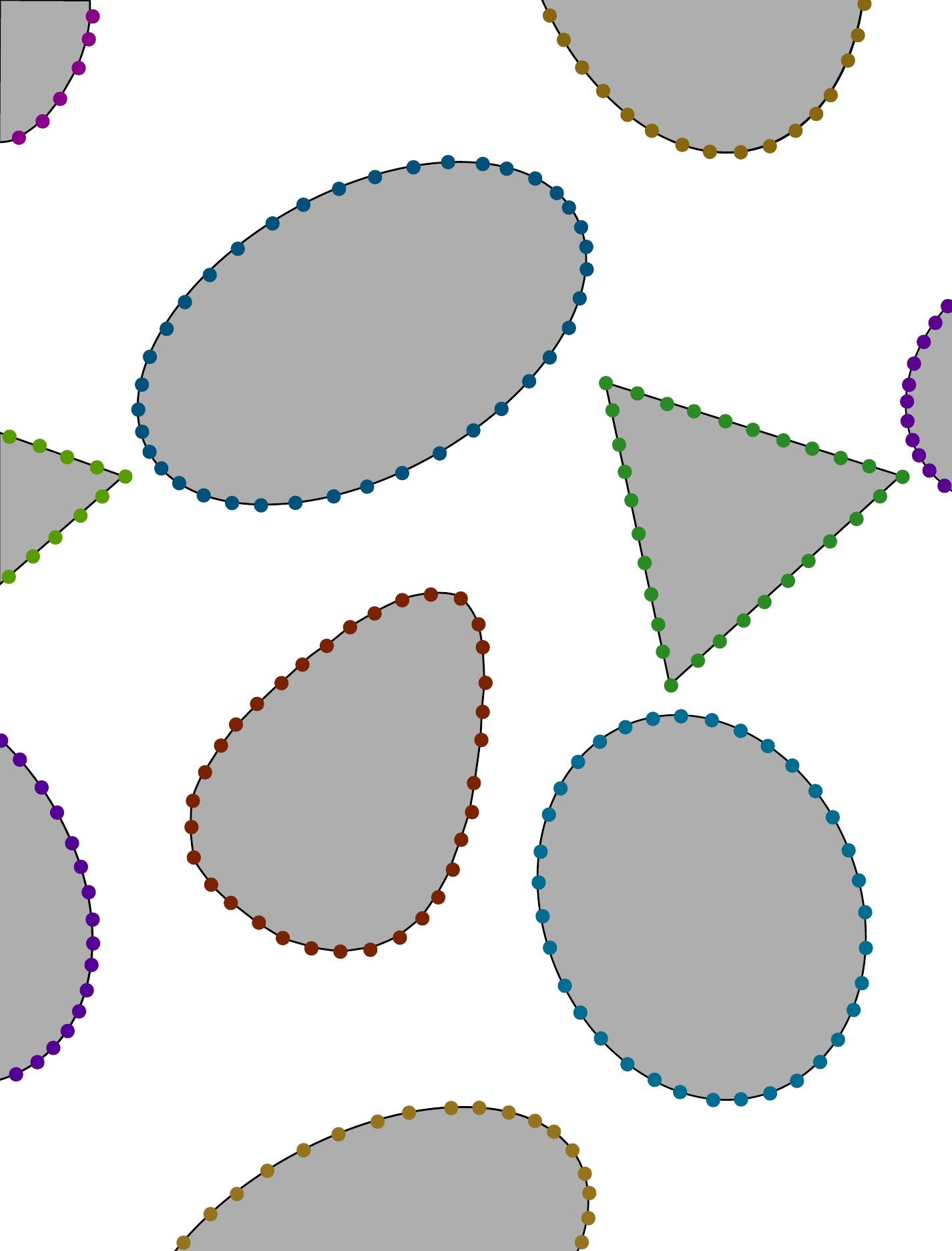
    \end{adjustbox}
    \hfill
    \begin{adjustbox}{minipage=0.19\textwidth}
    \def\svgwidth{\textwidth}
    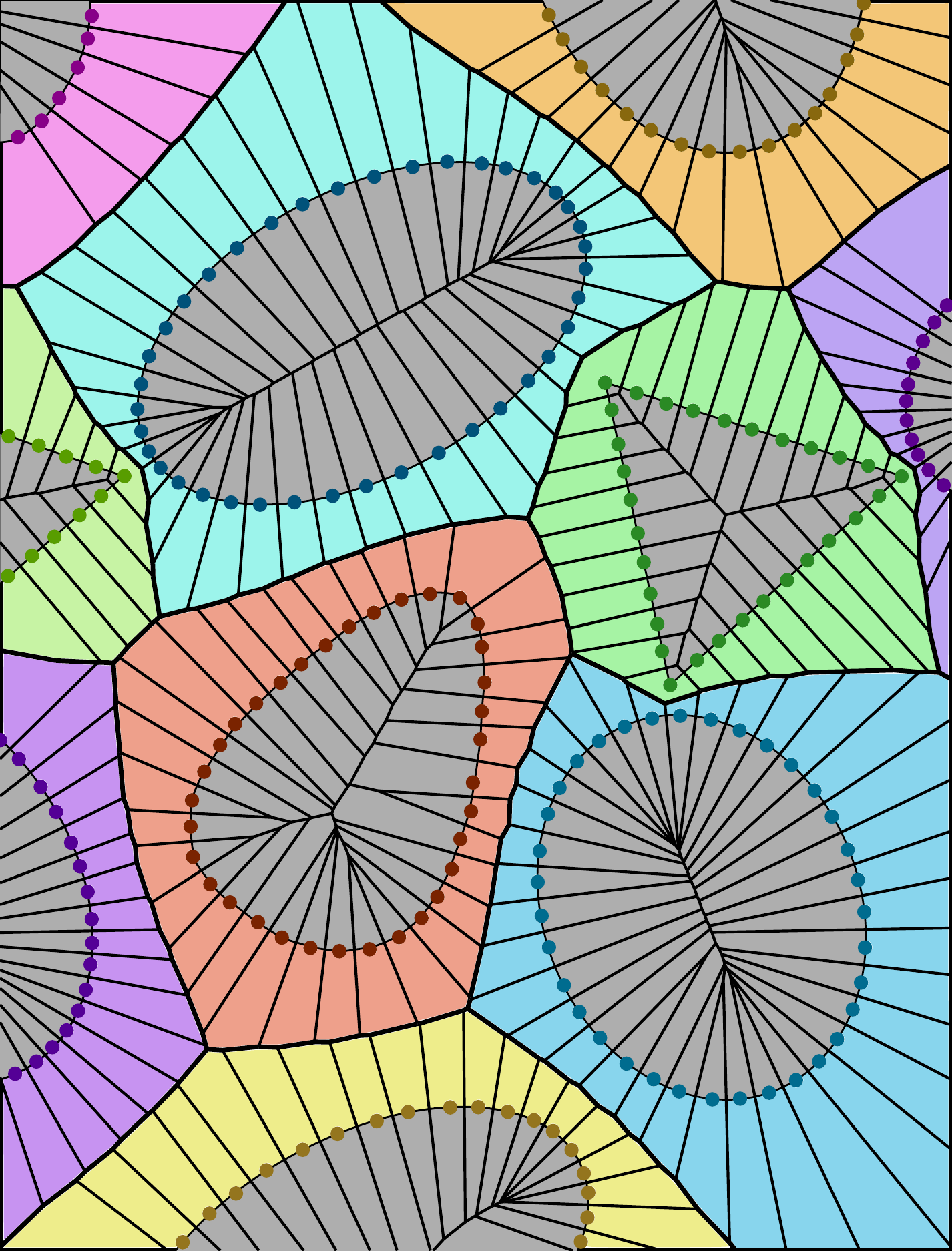
    \end{adjustbox}
    \hfill
    \begin{adjustbox}{minipage=0.19\textwidth}
    \def\svgwidth{\textwidth}
    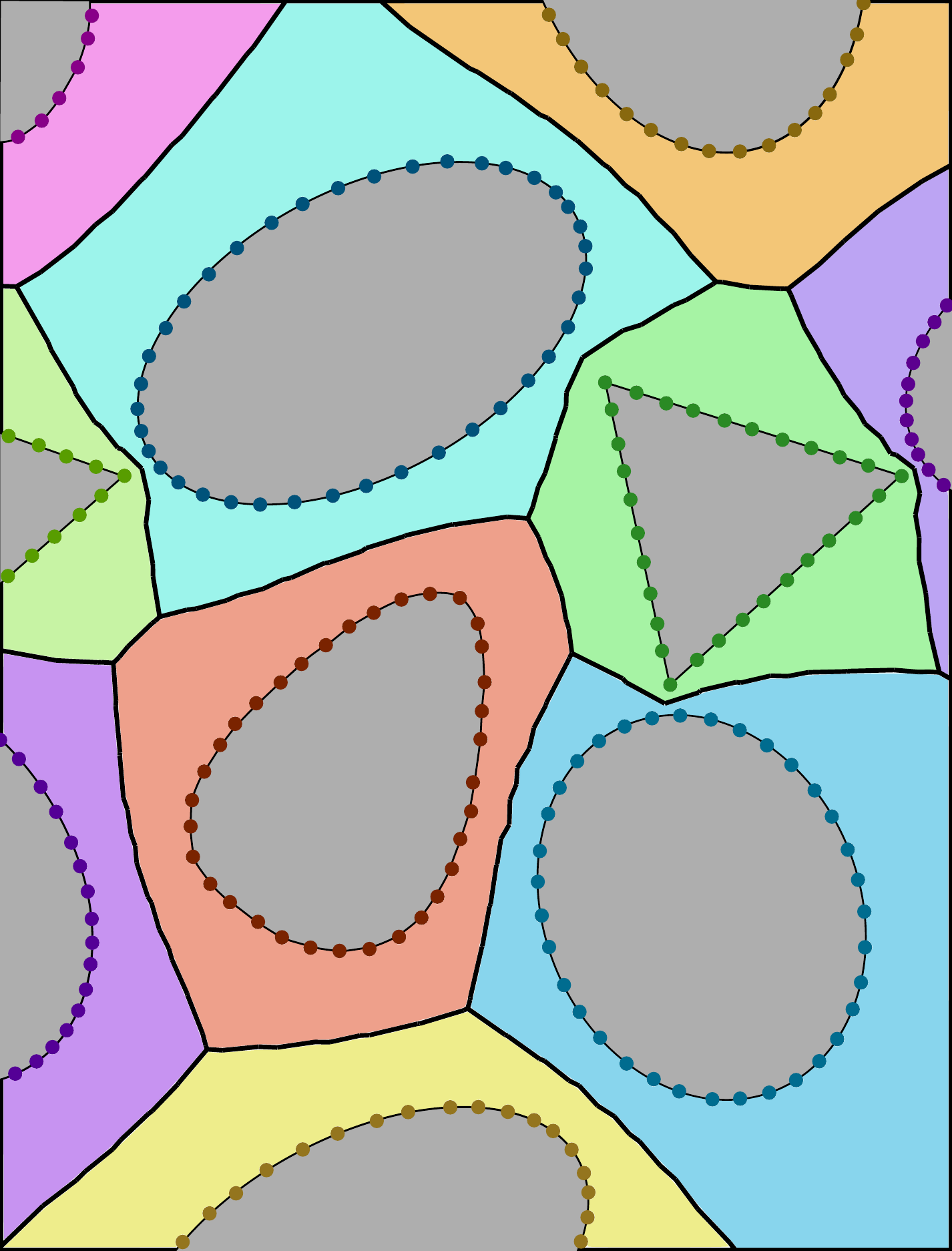
    \end{adjustbox}
    \hfill
    \caption{Sketch of the Set Voronoi algorithm from left to right. Generating points on particle surface. Calculating Voronoi Diagram of surface points. Merging cells belonging to the same particle.}
    \label{fig:creating_setvoronois}
\end{figure*}

The analysis of geometries and structures on a micro scale level is an important aspect of granular and soft matter physics to attain knowledge about many interesting properties of particle packings, including contact numbers, anisotropy, local volume fraction, etc. \cite{RefHecke,RefAste,RefWang}. 
A well-established concept is the so called \textit{Voronoi Diagram}. Here, the system is investigated by dividing the space into separate cells in respect to the positions of the center of the particles.
A cell assigned to a certain particle is defined as the space (or region of space) that contains all the volume closer to the center of this specific particle than to any other one (see figure\,\ref{fig:voronoi-setvoronoi} left). 
This partition of space, however, only yields precise results for monodisperse spheres as the construction fails otherwise due to morphological properties of the objects. 
For nonspherical or polydisperse particles the classical Voronoi diagram is of limited usefullnes, as shown in figure \ref{fig:voronoi-setvoronoi} (center) for a system of bidisperse spheres.
A generalized version of the Voronoi Diagram, the \textit{Set Voronoi Diagram} \cite{RefSchaller2013}, also known as navigational map \cite{RefLuchnikov} or tessellation by zone of influence \cite{RefPreteux}, has to be applied. 
In this case the cells contain all space around the particle which is closer to the particle's surface than to the surface of any other particle. 
Figure \ref{fig:voronoi-setvoronoi} (right) shows the Set Voronoi Diagram of a mixture of differently shaped particles.\\
Here we introduce a software tool called \textit{Pomelo}, which calculates Set Voronoi Diagrams based on the algorithm described in \cite{RefSchaller2013}.
Pomelo is particularly versatile due to its ability to generically handle any arbitrary shape which can be described mathematically. 
For instance objects neither have to be convex nor simply connected. 
In the first step of the algorithm (figure \ref{fig:creating_setvoronois} left) a triangulation of the particle's bounding surface is generated to sample its shape. 
\textit{Pomelo} offers functionality for some common particle shapes, but the user is can also specify generic particle shapes.
After the discretisation of the surface the system is tessellated by calculating the classical Voronoi diagram of all surface points of this triangulation. 
This is shown in figure \ref{fig:creating_setvoronois} (center).
To get the Set Voronoi tessellation of the system, cells belonging to points on the same particle surface are merged to a single cell in the last step (figure \ref{fig:creating_setvoronois} right). 
The resulting partition represents the Set Voronoi Diagram of the system. 
With this algorithm, systems and mixtures of particles of any arbitrary shape can be treated.

\section{Pomelo}
\label{sec:Pomelo}

\textit{Pomelo} is a generic Set Voronoi tool written in \texttt{c++11} and licensed under GPL3. 
\textit{Pomelo} can be downloaded, see ref. \cite{RefPomeloDownload}, as well as all instructions regarding setting up, building and using \textit{Pomelo}.
The system requirements are \texttt{g++ 4.9.2} or \texttt{clang++ 3.5.0-10} or any higher version.

While \textit{Pomelo}  can directly handle common particle shapes (mono- and polydisperse spheres, tetrahedra, ellipsoids and spherocylinders), it also provides a \texttt{generic} mode. 
The latter works for any shape which surface can be described mathematically.
The following two sections will describe how to use \textit{Pomelo} in both cases.
To use \textit{Pomelo} in \texttt{generic} mode (see section \ref{sec:generic_mode}), \texttt{lib-lua 5.2} or higher is required.

\subsection{Common Particle Shapes}
Particle shapes that are intrinsically supported by \textit{Pomelo} are mono- and polydisperse spheres, tetrahedra, spherocylinders and ellipsoids.
\textit{Pomelo} comes with a set of demos and tests.

One test case is a set of polydisperse spheres in a cubic cell. 
The input is a \texttt{xyzr} file. Its first line is the number of particles.
The second line is a comment which contains some information (separated by comma) required by \textit{Pomelo}. 
This includes boundary conditions (periodic/non periodic in each axis), box size, shrink and the number steps in $\phi$ and $\theta$ for discretizing the sphere's surface.
Every following line describes one sphere in the packing with its parameters (coordinates and radius).
To run the demo, call \textit{Pomelo} with the command line argument \texttt{-SPHEREPOLY}, the path to the \texttt{xyzr} file and the desired output folder.

\textit{Pomelo}'s output are the vertices, faces and cells of the Set Voronoi Diagram. The output can be set to different formats, like the POLY file format, off (for geomview) or a gnuplot readable format for easy visualisation.

\subsection{\texttt{Generic} Mode}
\label{sec:generic_mode}
Using \textit{Pomelo} in generic mode allows to calculate the Set Voronoi Diagram for generic particles. 
The input is a \texttt{position} file, which is a list of all particles which are described by a set of parameters each. 
The parameters are a complete description of the particles surface within the packing. 
The \texttt{read} file is key for \textit{Pomelo}s versatility. 
It allows the user to create a surface triangulation based on the parameters given in the \texttt{position} file with a script written in the lua language \cite{RefLua}. 
The script tells \textit{Pomelo} how to create a surface triangulation with the particles parameters as given in the \texttt{position} file. 
It can be fully customized by the user to match any specific particle shape.
This allows the user to create even systems composed of a mixture of diferent particles.
The surface triangulation of all particles will then be used to calculate the Set Voronoi Diagram as described above.

\textit{Pomelo}'s demos include a variety of examples on how to use the \texttt{read} script to handle different types of particles. 

\section{Applications of Set Voronoi Diagrams}
\label{sec:Applications}
\subsection{Ellipsoid Packings}

\begin{figure}[htbp]
    \centering
    \includegraphics[width=0.2\textwidth]{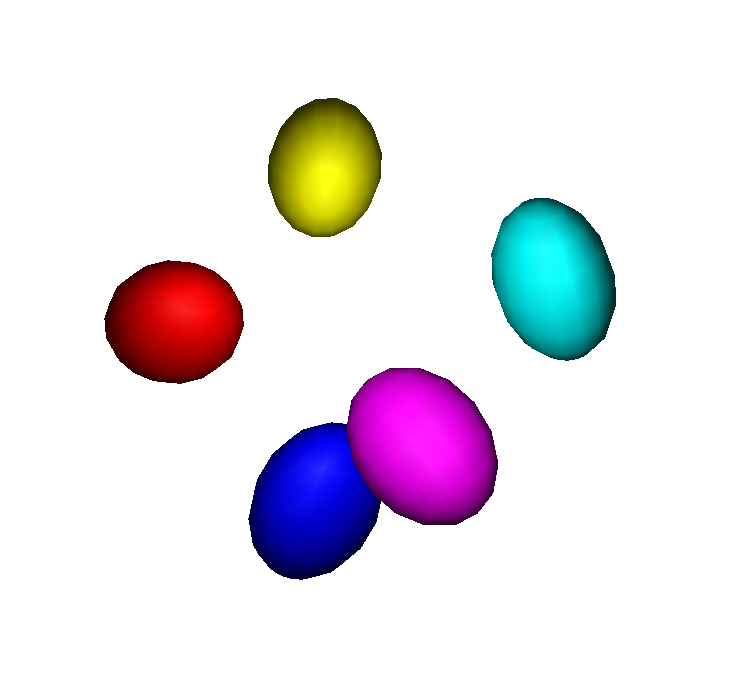}
    \includegraphics[width=0.2\textwidth]{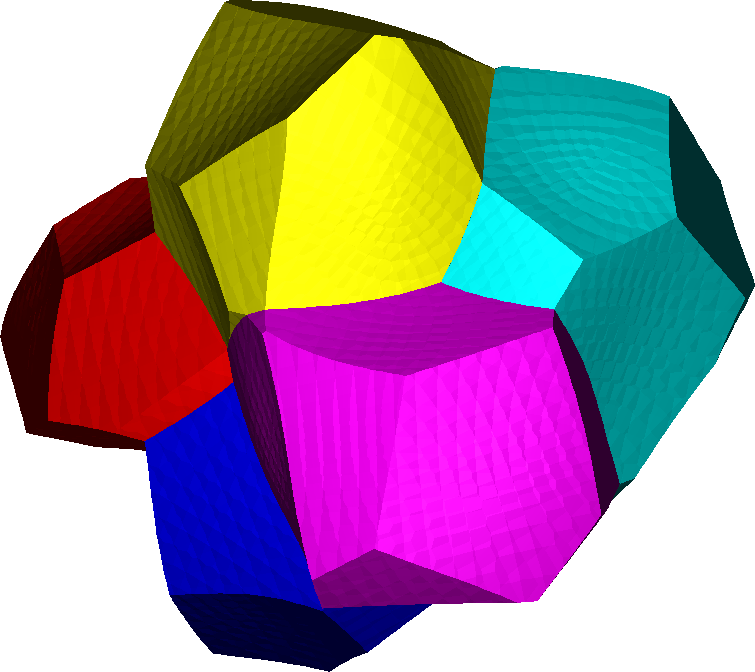}
    \caption{Ellipsoids ($\alpha=1.4$) in an isotopic system (left) and ther Set Voronoi Cells (right).}
    \label{fig:EllipVoronoi}
\end{figure}

To demonstrate the use of Set Voronoi Diagrams the local packing fraction which is defined as $\Phi_l = \frac{v_e}{v_i}$ with $v_i$ being the volume of the Set Voronoi Cell and $v_e = \frac{4}{3} \pi a b c$ the volume of the particle is calculated. It has been shown that the probability distribution of local packing fractions is universal for sphere packings \cite{RefAste} and jammed pro- and oblate ellipsoid packings \cite{RefSchaller2015}. 

\begin{figure}[htbp]
    \includegraphics{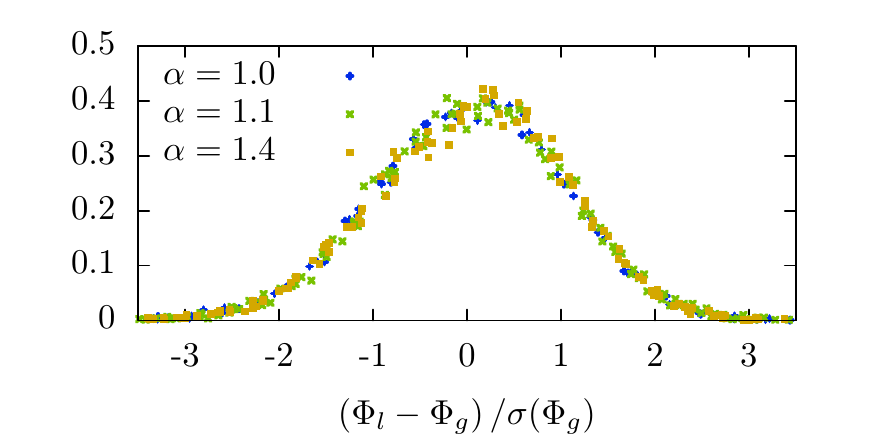}
    \caption{Probability distribution of rescaled local packing fractions $\Phi_l$ for packings of spheres (particle aspect ratio $\alpha=1.0$) and triaxial ellipsoids ($\alpha=1.1$, $\alpha=1.4$). 
    Global packing fractions range from 0.62 to 0.65.}
    \label{fig:ellipsoids_lpf}
\end{figure}

Experimental granular packings of triaxial ellipsoids (three distinct axis lengths $a$, $b$, $c$) have been measured at various global packing fractions using X-ray tomography, see figure \ref{fig:EllipVoronoi}.
Particle positions, orientations and size have been determined and the Set Voronoi Diagrams have been calculated for each packing. Here the particles are shrinked by \textit{Pomelo} to improve the quality of the cells. 
The statistical distribution of local packing fractions $\Phi_l$ for spheres ($\alpha =1.0$) and triaxial ellipsoids ($\alpha=1.1$, $\alpha=1.4$) is shown in figure \ref{fig:ellipsoids_lpf}.
To first order, the distribution can be collapsed on a master curve, as shown in Ref. \cite{RefSchaller2015}, by subtracting the global volume fraction $\Phi_g=\frac{<v_e>}{<v_i>}$ and scaling the distribution by $\sigma(\Phi_l)$, see figure \ref{fig:ellipsoids_lpf}. 
Thus the functional form of the distribution is invariant to global packing fraction $\Phi_g$ and aspect ratio $\alpha$, to first order.

\subsection{Tetrahedra}
Packings of tetrahedra can also be treated with \textit{Pomelo} without using the generic mode. 
The data was obtained by X-ray tomography \cite{RefNeudecker}.
Each tetrahedra is described by the positions of its 4 vertices.
\textit{Pomelo} is able to process the data directly. 
This gives the advantage to treat not only tetrahedra but pyramids in general.

\begin{figure}[htbp]
    \centering
    \includegraphics[width=0.2\textwidth]{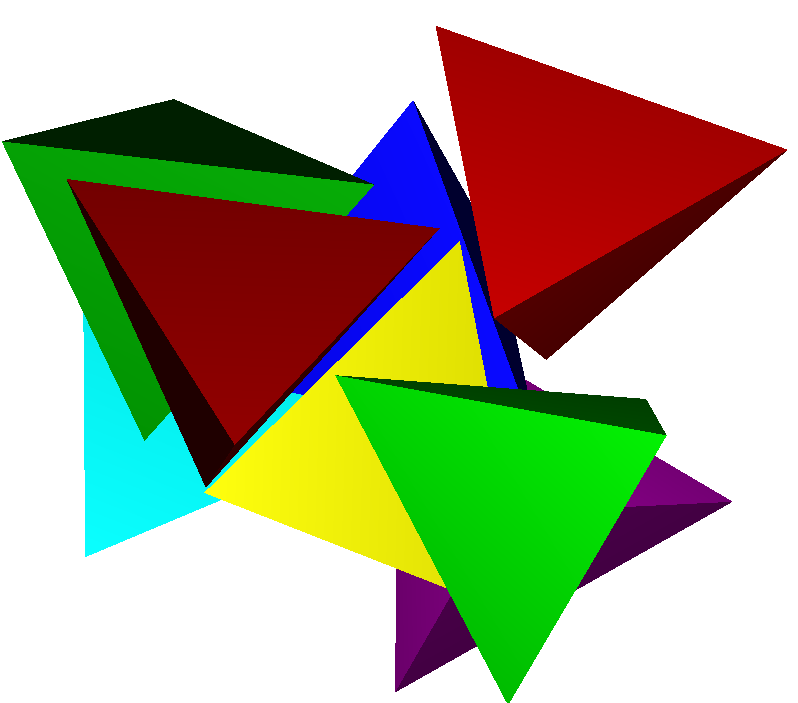}
    \includegraphics[width=0.2\textwidth]{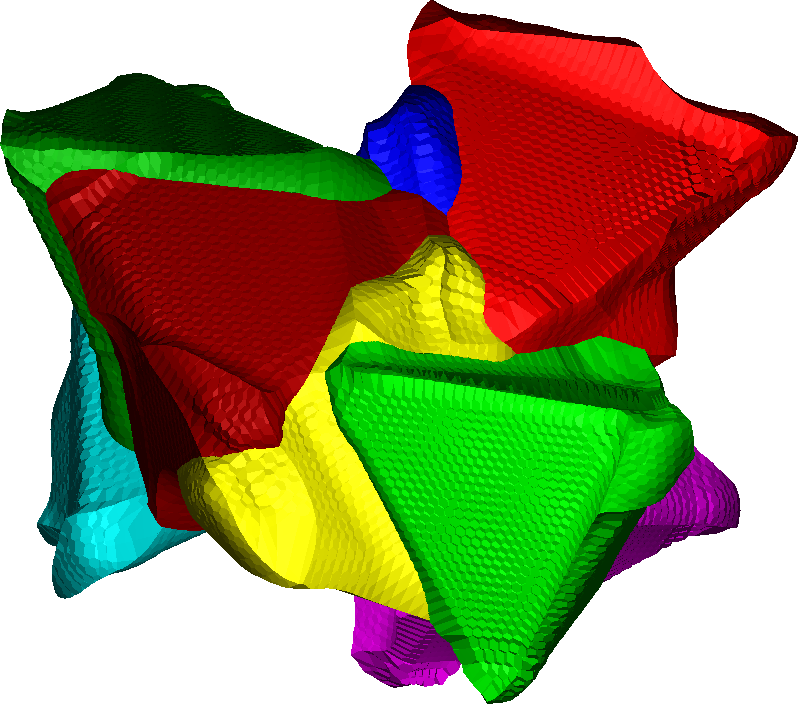}
    \caption{
        Tetrahedra of one of the experiments (left) and their Set Voronoi cels (right).
    }
    \label{fig:TetraVoronoi}
\end{figure}

\begin{figure}[htbp]
    \includegraphics{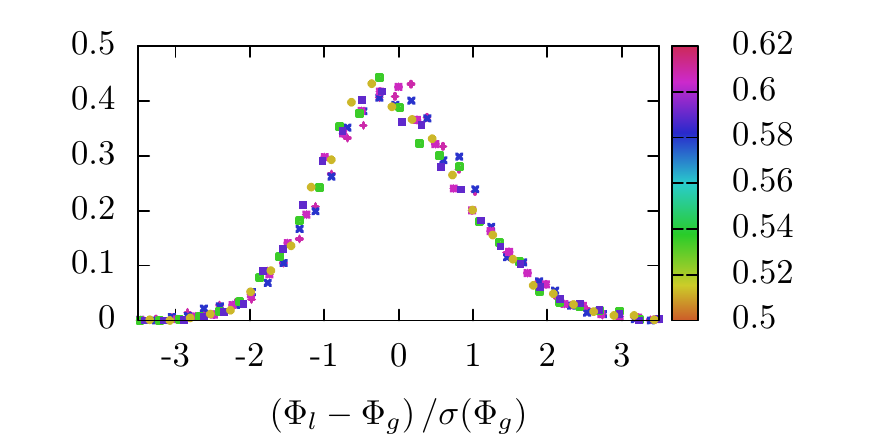}
    \caption{
    Probability distribution of rescaled local packing fractions $\Phi_l$ for different systems of tetrahedra.
    The colorbar and the point's color show the global packing fraction of the system.
}
    \label{fig:tetrahedra_lpf}
\end{figure}

The Set Voronoi tessellations of tetrahedra packings (figure \ref{fig:TetraVoronoi}) with different packing fraction and contact numbers are calculated with \textit{Pomelo}.
The local packing fraction is shown for those systems in figure \ref{fig:tetrahedra_lpf}.

\subsection{Pear Shaped Particles}

\begin{figure}[bp]
    \centering
    \includegraphics[width=0.26\textwidth]{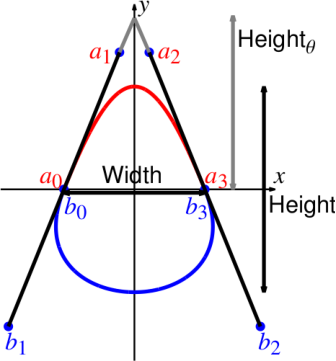}
    \caption{Pear shaped by two Bézier curves. Both are determined by the aspect ratio $\alpha=\frac{\text{Height}}{\text{Width}}$ and the degree of 
    tapering $\alpha_{\theta}=\frac{\text{Height}_{\theta}}{\text{Width}}$. }
    \label{fig:Bezier}
\end{figure}

The third example is the calculation of the Set Voronoi Diagrams of pear shaped or tapered particles. 
The shape of these particles is described by the aspect ratio $\alpha$ and the degree of tapering $\alpha_{\theta}$. 
By using two Bézier curves forming the bottom and top part of the pear and rotating them around their symmetry axis the surface of the particle is generated (figure\,\ref{fig:Bezier}) \cite{RefBarmes}. 
A triangulation algorithm of this surface is implemented within the \texttt{read} file, which allows \textit{Pomelo} to process pear shaped particles in the generic mode. 
The \texttt{position} file provides the position, orientation, size, aspect ratio and degree of tapering for every individual particle.

\begin{figure}[htbp]
    \centering
    \includegraphics[width=0.2\textwidth]{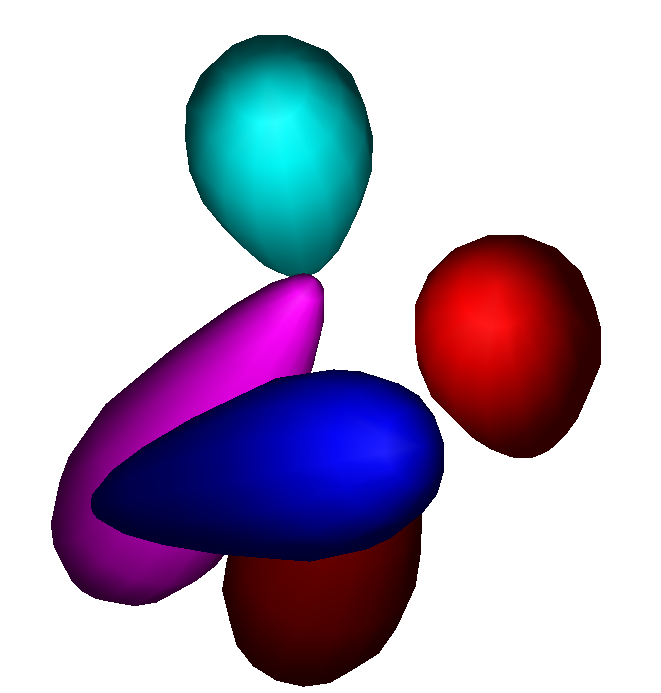}
    \includegraphics[width=0.2\textwidth]{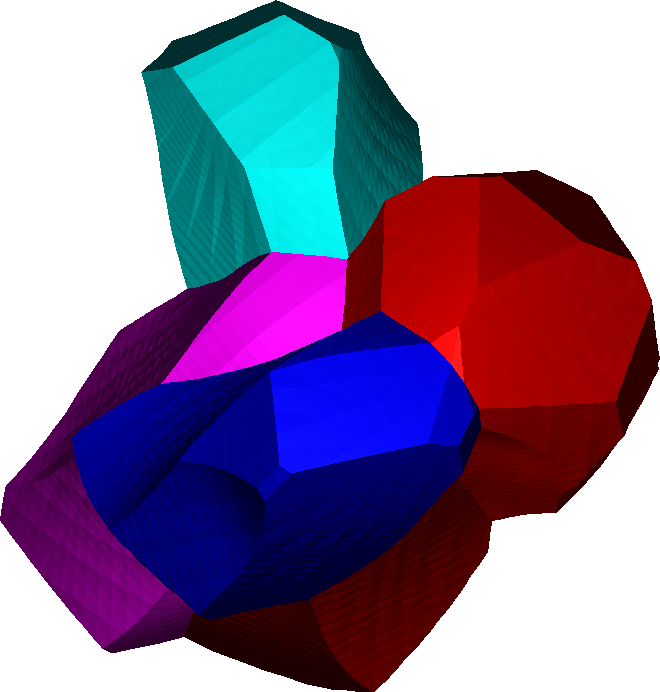}
    \caption{Pears ($\alpha=3.0, \alpha_{\theta}=3.8$) in an isotropic system (left) and their Set Voronoi cells (right).}
    \label{fig:PearVoronoi}
\end{figure}
\begin{figure}[htbp]
    \includegraphics{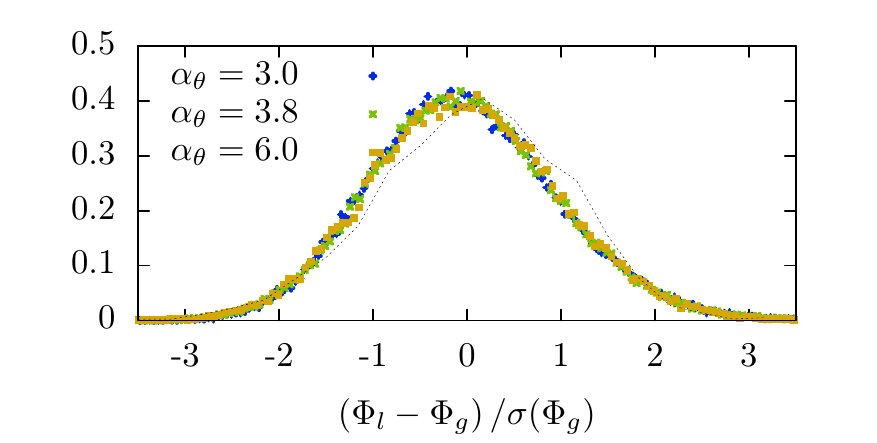}
    \caption{
        Probability distribution of rescaled local packing fractions $\Phi_l$ for pears ($\alpha=3.0$) with different degree of tapering and packing fraction $(\alpha_{\theta},\Phi_g)\in\{(3.0,0.48);(3.8,0.50);(6.0,0.54)\}$. 
        The dashed line shows the master curve of the sphere distributions.}
    \label{fig:pears_lpf}
\end{figure}

Systems of hard pear shaped particles ($\alpha=3.0$) with different degrees of tapering are generated using Molecular Dynamics (see figure \ref{fig:PearVoronoi}) \cite{RefBarmes}. 
The statistical distribution of the local packing fractions $\Phi_l$ for different pear systems ($\alpha_{\theta}=3.0,\Phi_g=0.48$ ; $\alpha_{\theta}=3.8,\Phi_g=0.50$ ; $\alpha_{\theta}=6.0,\Phi_g=0.54$) is shown in figure \ref{fig:pears_lpf}.
Similarly to the sphere and ellipsoid packings the distributions of the different pear systems collapse on a master curve. Accordingly the distribution is not only invariant to global packing fraction $\Phi_g$ but also to degree of tappering $\alpha_{\theta}$ for pears. However, the data of ellipsoids/spheres and pears do not collapse on the same curve.

\section{Outlook}
We have shown that \textit{Pomelo} is applicable for a variety of systems, including spheres, ellipsoids, tetrahedra and pear shaped particles.
We have illustrated its use to calculate Set Voronoi volume distributions. However, it is conceivable to analyse other interesting measures like Minkowski tensors, by piping the Set Voronoi Cells into corresponding analysis tools \cite{RefSchaller2015,RefSchroederTurk2011,RefSchroederTurk2013}. 
It is yet also unclear how similar volume distributions of different systems are. 

One way to improve the calculation of the Set Voronoi Diagram is to implement an adaptive sampling of the particle's surface. 
By comparing the distance between two neighboring surface points to the distance to the face of the Set Voronoi Cell it is possible to obtain an estimate on where the surface triangulation needs a better resolution and where a coarse resolution is good enough. 
This estimate will be used to change the density of the surface triangulation of the particles.

%
%
%

\end{document}